\documentclass[12pt,numbers]{article}

\usepackage[margin=2.5cm]{geometry}
\usepackage{soul}
\usepackage{braket, siunitx}
\usepackage{graphicx}
\usepackage{dcolumn}
\usepackage{bm}
\usepackage{braket, siunitx}
\usepackage{amssymb, amsmath}
\usepackage{multirow}
\usepackage[acronym]{glossaries}
\usepackage[autostyle]{csquotes}
\usepackage{algorithm}
\usepackage{algpseudocode}
\usepackage{setspace}
\usepackage{enumitem}
\usepackage{tabularx}

\usepackage[sort&compress,numbers]{natbib}
\usepackage{authblk}
\usepackage{xcolor}
\usepackage{booktabs}
\usepackage{tablefootnote}
\usepackage{footnote}
\usepackage[flushleft]{threeparttable}

\usepackage{algorithm}
\usepackage{algpseudocode}

\usepackage{doi}

\makeatletter
\def\@fnsymbol#1{\ensuremath{\ifcase#1\or \ddagger\or z\or \ddagger\or
\mathsection\or \mathparagraph\or \|\or **\or \dagger\dagger
\or \ddagger\ddagger \else\@ctrerr\fi}}
\makeatother

\newcommand*{\od}[3][]{\relax\ifmmode{\dfrac{\mathrm{d}^{#1} #2}{\mathrm{d} #3^{#1}}}\else{${\mathrm{d}^{#1} #2}/{\mathrm{d} #3^{#1}}$}\fi}
\newcommand*{\pd}[3][]{\relax\ifmmode{\dfrac{\partial^{#1} #2}{\partial #3^{#1}}}\else{${\partial^{#1} #2}/{\partial #3^{#1}}$}\fi}

\newacronym{ev}{EVs}{electric vehicles}
\newacronym{lib}{LIBs}{lithium-ion batteries}
\newacronym{bms}{BMS}{battery management system}
\newacronym{rnn}{RNN}{recurrent neural network}
\newacronym{lstm}{LSTM}{long short-term memory}
\newacronym{ctcv}{CtCV}{cell-to-cell variability}
\newacronym{phm}{PHM}{prognostics and health management}
\newacronym{ml}{ML}{machine learning}
\newacronym{shap}{SHAP}{Shapley additive explanation}
\newacronym{eol}{EoL}{end-of-life}
\newacronym{snn}{SNN}{Sequential neural network}
\newacronym{pcc}{PCC}{Pearson correlation coefficient}
\newacronym{cc}{CC}{constant current}
\newacronym{cv}{CV}{constant voltege}
\newacronym{soh}{SOH}{state-of-health}
\newacronym{rmse}{RMSE}{root mean square error}

\providecommand{\keywords}[1]
{
  \small	
  \textbf{\textit{Keywords---}} #1
}

\title{State-of-Health Prediction for EV Lithium-Ion Batteries via DLinear and Robust Explainable Feature Selection}

\author{ Minsu Kim$^{a,1}$, Jaehyun Oh$^{b,1}$, Sang-Young Lee$^{b,c,}$\thanks{Corresponding author \texttt{syleek@yonsei.ac.kr}.}, Junghwan Kim$^{b,c,}$\thanks{Corresponding author \texttt{kjh24@yonsei.ac.kr}.} \\
\small $^{a}$Department of Chemical Engineering, Massachusetts Institute of Technology, Cambridge, Massachusetts 02139, United States of America\\
\small $^{b}$Department of Chemical and Biomolecular Engineering, Yonsei University, Seoul 03722, Republic of Korea\\
\small $^{c}$Department of Battery Engineering, Yonsei University, Seoul 03722, Republic of Korea}

\begin{document}

\maketitle

\footnotetext[1]{These authors contributed equally to this work.}

\begin{abstract}
Accurate prediction of the state-of-health (SOH) of lithium-ion batteries is essential for ensuring the safety, reliability, and efficient operation of electric vehicles (EVs). Battery packs in EVs experience nonuniform degradation due to cell-to-cell variability (CtCV), posing a major challenge for real-time battery management. In this work, we propose an explainable, data-driven SOH prediction framework tailored for EV battery management systems (BMS). The approach combines robust feature engineering with a DLinear. Using NASA's battery aging dataset, we extract twenty meaningful features from voltage, current, temperature, and time profiles, and select key features using Pearson correlation and Shapley additive explanations (SHAP). The SHAP-based selection yields consistent feature importance across multiple cells, effectively capturing CtCV. The DLinear algorithm outperforms long short-term memory (LSTM) and Transformer architectures in prediction accuracy, while requiring fewer training cycles and lower computational cost. This work offers a scalable and interpretable framework for SOH forecasting, enabling practical implementation in EV BMS and promoting safer, more efficient electric mobility.
\end{abstract}


\newpage
\doublespacing
\section{Introduction}

{T}he \gls{ev} industry has grown rapidly due to environmental issues related to internal combustion engine vehicles. Lithium-ion batteries (LIBs) are the most widely used battery materials based on their advantages such as high energy density, low self-discharge characteristics, and long lifespan \cite{dunn2011electrical, li2024recovery, seol2025recent}. However, it is inevitable that battery performance degrades as impedance increases due to chemical decomposition and lithium plating during battery usage. Monitoring capacity reduction and accurately predicting \gls{soh} through \gls{bms} have become crucial, as the performance degradation of LIBs can pose critical risks, such as thermal runaway reactions \cite{finegan2017runaway, mao2018runaway, shin2023feature}. Predicting the necessary replacement time in advance is challenging, due to intra-cell factors including morphology, architecture, and composition, as well as inter-cell factors such as winding, connection, and assembly processes that contribute to \gls{ctcv} \cite{kim2024advanced, kim2024fast}, along with differences in operating conditions for each device. The \gls{phm} system implements condition-based management by predicting the necessary replacement chance before failures occur, allowing for repairment only when it’s necessary \cite{goebel2008phm, beck2021ctcv}. It considers battery capacity degradation using various sensor data, such as voltage, current, and temperature.
\gls{soh} prediction is one of the critical functions of the \gls{phm} system, which forecasts the remaining cycles where the battery can be used safely. Accurate \gls{soh} prediction contributes to optimizing battery replacement timing and prevents accidents caused by excessive charging or discharging.

\gls{soh} prediction methods using early cycling behavior can generally be classified into two approaches: model-based approach and data-driven approach \cite{prasad2013aging, jiangtao2020vcmodel}. However, model-based approaches often encounter challenges. In practice, the irreversible capacity fade is caused by an interaction of mechanical (e.g., particle cracking) and chemical (e.g., electrolyte decomposition) factors, making it quite difficult to mathematically identify these degradation mechanisms. The existence of a knee point at which aging accelerates in a certain cycle, and often a slight capacity increase early in the cycle, makes model-based approaches difficult \cite{severson2019data}. Furthermore, while battery models such as the Doyle-Fuller-Newman (DFN) \cite{doyle1993modeling, newman1975porous} model have been successfully applied to predict \gls{soh}, the strong nonlinearity of these models necessitates further analysis, including identifiability analysis \cite{ramadesigan2011parameter,berliner2025bayesian}. The data-driven approach is a promising alternative to overcome this challenge \cite{zhou2020attention, severson2019data, ai2022mftransformer}.

The data-driven \gls{soh} prediction research typically employs time-series \gls{ml} algorithms, with changes observed for each charging/discharging cycle. Normally, \gls{rnn} \cite{hopfield1982rnn,errorpropa1986Rumelhart,jordan1986serial} and \gls{lstm} \cite{jurgen1997lstm} have been used for such tasks: Liu et al. \cite{liu2010adaptive} propose an adaptive \gls{rnn} that continuously updates the parameters of the network to improve the description performance of dynamic cycling behavior of the LIBs. Park et al. \cite{park2020lstm} conducted \gls{soh} forecasting using \gls{lstm}, but they struggle with learning long-term dependency \cite{bengio1994longdependency}. To overcome this limitation, the attention algorithm was introduced, enabling predictions that consider a broader context rather than just neighboring time steps \cite{bahdanau2016neural, luong2015nmtattention}. Designed for natural language processing tasks and centered around the attention algorithm, Transformer architectures outperform \gls{rnn} and \gls{lstm} by taking a holistic view of time-series data to learn features of data, then trains in order of importance rather than sequential dependency \cite{google2017transformer}. Chen et al. \cite{chen2022transformer} conducted a prediction of \gls{soh} by Transformer, successfully overcoming the long-term dependency hurdle of RNN and \gls{lstm}.  Despite being a game-changer in time series forecasting, the Transformer has encountered difficulties when applied to time series prediction \cite{zeng2023dlinear}. Specifically, its high complexity in space (i.e., memory) time and its performance are not significantly superior to conventional \gls{lstm} models. To address these challenges, DLinear was proposed \cite{zeng2023dlinear}. DLinear uses two linear layers considering trend and seasonality, in contrast to the complex structure and extensive computation of the Transformer. Unlike the Transformer, DLinear takes both trend and seasonality into account in time series data predictions. Furthermore, it provides explainablity with its weights of linear layers, revealing parameter importance. The SOH of the battery typically exhibits a decreasing pattern during its lifespan, with capacity increases occasionally observed at the early cycles, punctuated by periodic increases. Although the Transformer finds it challenging to grasp this trend, DLinear effectively captures both the downward and occasional upward movements, surpassing LSTM and Transformer-based predictions in terms of both space and time complexity.

The performance of data-driven \gls{soh} prediction, such as DLinear, is determined by the quantity and quality of data. Since battery cycling experiments require a significant budget in time and cost, there are practical difficulties in generating large-scale datasets. Therefore, it is essential to identify the key features that significantly contribute to battery health and to attain high predictive performance, even with a limited number of training cycles. However, because high-dimensional features can overfit the cycling behavior used for training, selecting key features is essential, and various feature engineering studies have been conducted to predict battery lifespan: Fei et al. \cite{fei2021early} extracted 42 features from a large-scale lithium iron phosphate (LFP) battery dataset \cite{severson2019data} and identified a high-performance prediction model structure by applying various feature selection approaches such as wrapper and filter approaches and various prediction algorithm. Hu et al. \cite{hu2020battery} proposed a fusion approach that combines the advantages of wrapper and filter approaches to select features that contribute to \gls{soh} prediction. Greenbank et al. \cite{greenbank2021automated} propose an automatic feature selection method using the \gls{pcc}, a filter approach, to predict the knee point where battery life declines rapidly and input the \gls{eol} prediction. However, the most widely used filter approach has concerns that correlation ranking can only consider linear dependencies between features and output. \gls{shap} \cite{lundberg2017unified}, which has recently been used in sensitivity analysis, provides insight at the feature selection stage because it presents qualitative results as well as quantitative sensitivity \cite{owen2014sobol, vuillod2023comparison}. Additionally, in Ref. \cite{gebreyesus2023machine}, \gls{shap} value-assisted feature selection was found to have superior performance compared to other algorithms for time series applications such as battery lifespan prediction. Moreover, the \gls{shap}-based approach, which is a model-agnostic approach, is useful when applied to various data-driven models. Based on the above considerations, this work focuses on feature engineering based on battery chemistry and data-driven prediction of \gls{soh}. Through sensitivity analysis, key features identified are used as inputs to a time-series forecasting model to accurately predict the decreasing trend of \gls{soh}. Specifically, our main contributions are presented below:
\begin{enumerate}[label=\arabic*)] 
    \item Using the data collected from the cycling experiment, a feature set required for \gls{soh} prediction is generated. \gls{pcc} and \gls{shap}-based global sensitivity analysis are performed to identify key features among all features.
    \item The key feature subset is used as input to \gls{lstm}, Transformer, and DLinear, and the accuracy is compared with previous works.
    \item The variability of cells composing modules or packs is intensively analyzed. Key features that are equally applied to the cells within specific modules or packs provide opportunities for efficient \gls{soh} prediction in practical \gls{bms}.
\end{enumerate}

This article is organized as follows: In Section \ref{sec2}, the three machine learning algorithms used for the prediction of \gls{soh} are described. In Section \ref{sec3}, the dataset used in the cycling experiment and the scheme for extracting features contributing to the battery lifespan based on battery chemistry are explained. Additionally, two algorithms used for feature selection are introduced. Section \ref{sec4} discusses feature engineering and \gls{soh} prediction results, and Section \ref{sec5} presents a summary of the results and future work.

\section{Time-series prediction algorithm} \label{sec2}

\begin{figure*}
    \centering
    \includegraphics[width=6in]{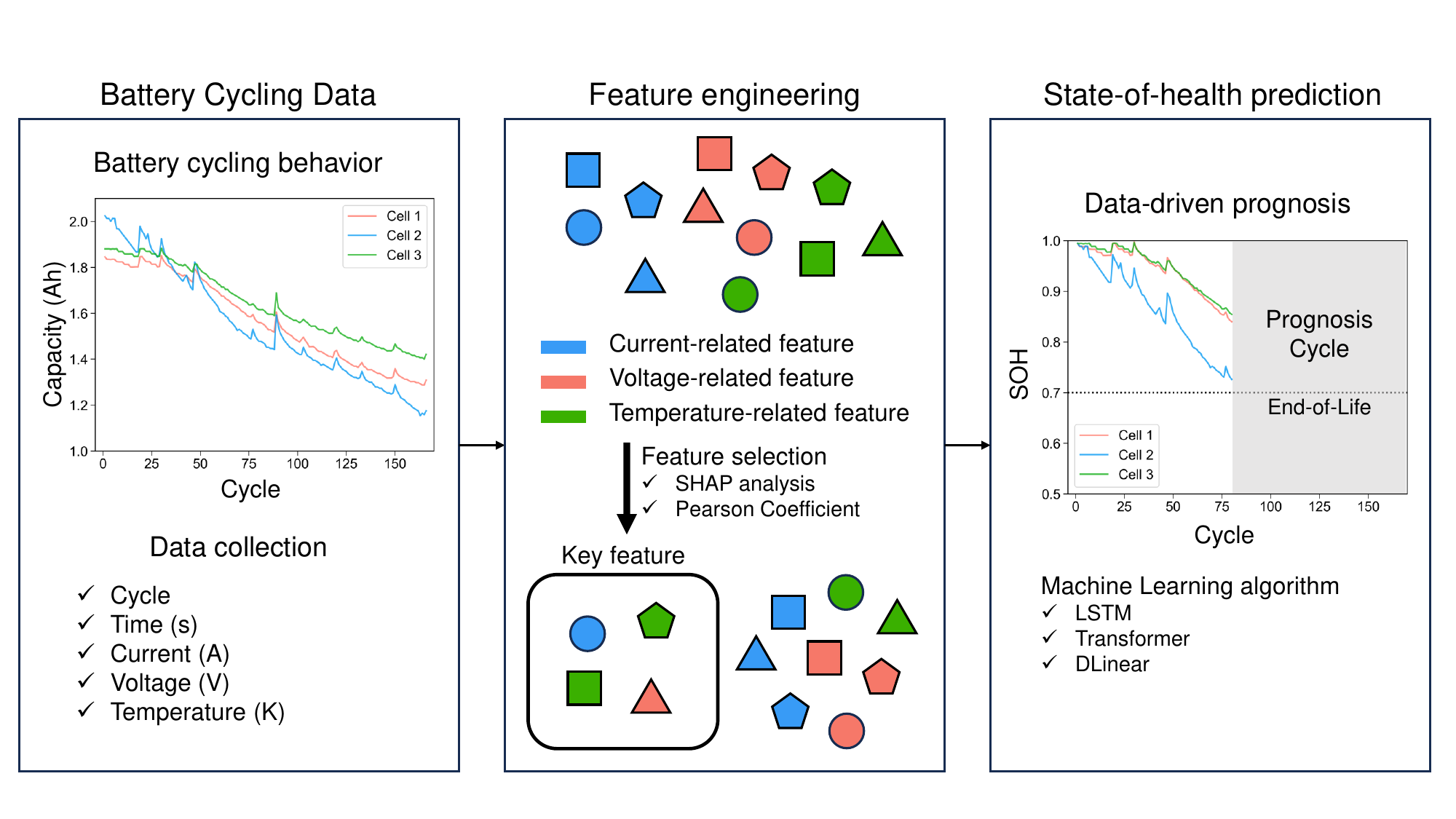}
    \caption{Schematic diagram of \gls{soh} prediction through feature engineering}
\label{fig1}
\end{figure*}

\subsection{Long short-term memory (LSTM)}
\gls{snn} process time-series data by assuming that the state at any given time step depends on prior time steps. \gls{rnn} is the pioneering \gls{snn} structure designed for processing time-series data \cite{hopfield1982rnn,errorpropa1986Rumelhart,jordan1986serial}. The \gls{rnn} algorithm takes input to produce output, and also sends its output back to itself, to recurrent into the next time step. Due to this structure, each neuron in \gls{rnn} has two sets of weights: one for input and another for multiplying with the output from the previous time step. In this way, \gls{rnn} takes previous information into account and determines the extent of this reference with weights. However, it has pointed that as time progresses, the continuous multiplication of weights in \gls{rnn} results in the dilution of information from distant past time steps at the current time step, a problem known as gradient vanishing \cite{Bengio1994Vanishing}.

To address this issue, the \gls{snn} known as \gls{lstm} was suggested, which divides the output from the previous time step into long-term state and short-term state. \gls{lstm} forwards them to the subsequent time steps. By introducing long-term state with minimal loss according to weights, \gls{lstm} improves the gradient vanishing problem observed in \gls{rnn}.

\subsection{Transformer}
For a long time, \gls{lstm} has been firmly established as the state-of-the-art approach in time series data. However, long-term state of \gls{lstm} also did not completely solve the gradient vanishing problem \cite{bengio1994longdependency}. The problem was alleviated but similar, as the consideration of information from distant past time steps being diluted as weights were multiplied with the long-term state of \gls{lstm} with the progression of time steps. Over time, an Attention algorithm was devised to focus more on contextually significant information after receiving information, not only from the immediate previous time step but from all time steps \cite{bahdanau2016neural, luong2015nmtattention}. This allowed for learning based on meaningful information from the entire dataset rather than focusing solely on nearby information, thereby addressing the issues of \gls{rnn}, \gls{lstm}, and similar SNN algorithms. Furthermore, implemented with the Attention algorithm, the Transformer was devised. Recently, Transformer-based approaches and variant models of Transformer were utilized in predicting the \gls{soh} of LIBs \cite{chen2022transformer}.

\subsection{DLinear}
The Transformer has successfully overcome the hurdle of gradient vanishing and long-range dependencies posed by \gls{lstm}. However, as the Transformer was designed for natural language processing tasks (e.g., translation, summarization), not for time-series data, it had the weakness of poor performance when applied to time-series data. In particular, early Transformer models had difficulty in identifying trends in data that increased or decreased continuously throughout the observation period. Furthermore, due to its complex structure for analyzing sentences and texts written in languages, the Transformer required excessive computation compared to the conventional SNNs, thereby demanding a significant amount of time and memory. The DLinear model was introduced to mitigate these two problems \cite{zeng2023dlinear}. Its inherently low computational burden, stemming from the linear structure of the model, offers a distinct advantage for \gls{bms}, as these systems necessitate real-time forecasting and are often constrained by the practical limitations of deploying models with excessive computational demands.

\begin{algorithm}
\caption{Decomposition-Linear (DLinear) Algorithm for \gls{soh} Prediction}
\begin{algorithmic}[1]
\State \textbf{Input:} Feature matrix $X: [F_1, F_2, ..., F_{20}]$, Target $y = \text{SoH}$

\State \textbf{Step 1:} Compute the moving average $X_t$ using a sliding window

\State \textbf{Step 2:} Decompose into trend ($X_t$) and remainder ($X_s$) with $X_s = X - X_t$

\State \textbf{Step 3:} Pass $X_t$ and $X_s$ through linear layers
    \If {$individual$ is True}
        \For{each feature $i$ in $X$}
            \State Apply linear layers to predict future trend and seasonal components
        \EndFor
    \Else
        \State Use shared linear layers for all features
    \EndIf

\State \textbf{Step 4:} Combine predicted $X_t$ and $X_s$ to get final prediction

\State \textbf{Output:} Predicted time series data

\end{algorithmic}
\end{algorithm}

The detailed procedure for \gls{soh} prediction using DLinear is introduced in \textit{Algorithm 1}. DLinear separates time-series data into Trend ($X_t$) and Remainder ($X_s$) using a moving average, and learns each through a separate linear layer. Equation of $X_t$ and $X_s$ can be defined as follows:
\begin{equation}
X_t = \frac{1}{n} \sum_{k=0}^{n-1} X_{i-k}
\end{equation}

\begin{equation}
X_s = X - X_t
\end{equation}
where $i$ is the time step of data, and $n$ is the window size of the moving average. Based on the learning of these two linear layers, the future Trend and Remainder are predicted, and these two predictions are combined to derive the final prediction.

\section{Feature engineering} \label{sec3} 
In this section, feature engineering to extract parameters contributing to \gls{soh} from cycling data of LIBs and identify key parameters is explained. Our \gls{soh} prediction framework is validated using the LIBs aging dataset provided by NASA's Prognostics Center of Excellence \cite{NasaData}. The nominal capacity of the four selected cells (i.e., B0005, B0006, B0007, and B0018) is 2 Ah. All the cells are cycled under the same charge and discharge procedure. In the charge phase, the cells are charged with a \gls{cc} of 1.5A until the voltage reaches 4.2V. After the voltage reaches 4.2V, they are charged with a \gls{cv} until the current reaches 0.02A. In the discharge phase, the cells are discharged in \gls{cc} mode with 2.0 A until B0005, B0006, B0007, and B0018 reach 2.7V, 2.5V, 2.2V, and 2.5V, respectively. In Section \ref{sec3_1}, the process of extracting features expected to contribute to lifespan through information on voltage, temperature, and time for four \gls{lib} cells is described. In Section \ref{sec3_2}, the process by which a significantly contributing subset of the entire feature set is selected through the filter approach and \gls{shap}-based approach is explained.

\subsection{Feature extraction} \label{sec3_1} 

\begin{figure}[ht!]
    \centering
    \includegraphics[width=14cm]{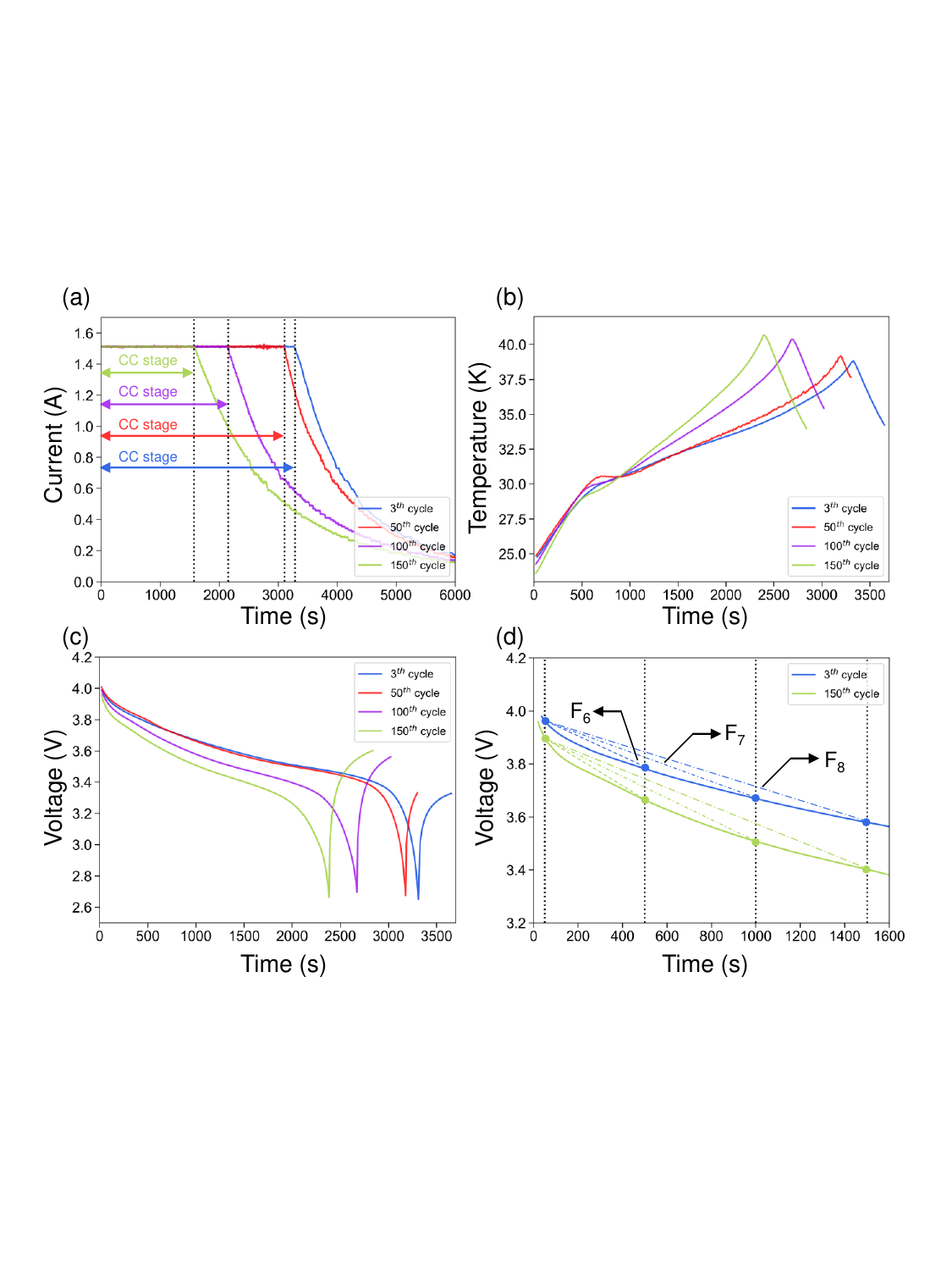} 
    \vspace{-4cm} 
    \caption{Feature description of charging and discharging for B0005 battery cell: (a) CC charging time (F$_{18}$), (b) Temperature in discharging, (c) Voltage in discharging, (d) Slope of discharge voltage curve (F$_{6}$, F$_{7}$, and F$_{8}$)}
\label{fig2}
\end{figure}

In feature extraction, factors that contribute to \gls{soh} prediction are extracted based on battery cycling data collected from the \gls{bms}. According to battery chemistry, the main factors that accelerate degradation are voltage \cite{klein2011optimal} and temperature \cite{leng2015effect} during cycling, and degradation causes differences in charge and discharge profiles. Consequently, even if a feature that cannot be explained by battery chemistry is identified as a key feature \cite{severson2019data}, it is still necessary to find key features that can explain a battery's health for \gls{soh} estimation. Table \ref{featureTable} describes 20 features generated from current, voltage, temperature, and time data collected in each cycle. Fig. \ref{fig2}(a-c) shows that the current and voltage profile at the charging stage, temperature distribution, and discharge voltage curve are dependent on the \gls{soh}. These dependent characteristics are processed through statistical information, such as skewness, and included in the feature set. In particular, the characteristic of the discharge curve, where differences occur after significant cycles, is explained by the slope at the early stage of the discharge (Fig. \ref{fig2}(d)).

\begin{table*}[!ht]
\centering
\caption{\\Description of features extracted for each parameter}
\label{featureTable}
\resizebox{5in}{!}{%
\begin{tabular}{ccc}
\hline
Related parameter   & Feature designation & Description                                     \\ \hline
Current and Voltage & F$_{1}$                  & Variance of measured current in discharging     \\
                    & F$_{2}$                  & Variance of measured voltage in discharging     \\
                    & F$_{3}$                  & Median of loaded voltage in discharging         \\
                    & F$_{4}$                  & Skewness of measured voltage in discharging     \\
                    & F$_{5}$                  & Skewness of loaded voltage in discharging       \\
                    & F$_{6}$                  & Slope of discharge voltage curve (50s to 500s)  \\
                    & F$_{7}$                  & Slope of discharge voltage curve (50s to 1000s) \\
                    & F$_{8}$                  & Slope of discharge voltage curve (50s to 1500s) \\ \hline
Temperature         & F$_{9}$                  & Maximum temperature in discharging              \\
                    & F$_{10}$                 & Average temperature in discharging              \\
                    & F$_{11}$                 & Variance of temperature in discharging          \\
                    & F$_{12}$                 & Skewness of temperature in discharging          \\
                    & F$_{13}$                 & Minimum temperature in discharging              \\
                    & F$_{14}$                 & Maximum temperature in charging                 \\
                    & F$_{15}$                 & Minimum temperature in charging                 \\
                    & F$_{16}$                 & Average temperature in charging                 \\
                    & F$_{17}$                 & Skewness of temperature in charging             \\ \hline
Time                & F$_{18}$                 & CC charging time                                \\
                    & F$_{19}$                 & CV charging time                                \\
                    & F$_{20}$                 & Total discharging time                          \\ \hline
\end{tabular}
}
\end{table*}

\newpage
\subsection{Feature selection} \label{sec3_2}
\subsubsection{Filter approach} 

The filter approach uses feature ranking to select key features. In general, the correlation between the extracted features and the objective function (i.e., \gls{soh}) scores the ranking of each feature, selects features that are higher than a certain threshold, and screens other features \cite{sedgwick2012pearson}. Because the filter approach is applied without being integrated into the prediction model, it can be easily applied to high-dimensional feature sets with relatively low computational budgets. This quantifies the linear correlation between the feature denoted by $x$ and the objective function denoted by $y$.

\begin{equation}
\label{filter_eq1}
r_{xy}=\frac{\sum_{i=1}^{n}(x_{i}-\bar{x})(y_{i}-\bar{y})}{\sqrt{\sum_{i=1}^{n}(x_{i}-\bar{x})^{2}\sum_{i=1}^{n}(y_{i}-\bar{y})^{2}}}.
\end{equation}

Where $\bar{x}$ and $\bar{y}$ are sample mean of $x$ and $y$, respectively. As for the key feature selection, all correlation coefficients between key features and cell capacity are evaluated based on Eq.~\eqref{filter_eq1}. The three parameter subsets with the highest absolute correlation coefficients are regarded as the \textit{most sensitive features}, while those with the lowest absolute values are considered the \textit{least sensitive features}. This concept also applies to the SHAP-based approach described in Section \ref{sec3_2_2_shap}

\subsubsection{SHAP-based approach} \label{sec3_2_2_shap}

\gls{shap} is an approach derived from Shapley values to explain the contribution of each feature to the predictions of a machine learning model. Originating from game theory, Shapley values aim to fairly distribute the contribution when players collectively achieve a specific outcome. Owen et al. \cite{owen2014sobol} identified the relationship between Shapley values and variance-based global sensitivity analysis. The sum of the Shapley values for all features used in training a machine learning model is exactly equal to the total variance. This normalization property is quite crucial when determining the importance ranking of features. Sobol' indices, widely used in global sensitivity analysis, provide contributions for individual features similarly to \gls{shap}, but when dependencies exist between features, the sum of sensitivities may not match the total variance \cite{li2022extracting}. In contrast, Shapley values work well even in the presence of dependencies between features. The Shapley value for feature F$_{i}$ of a machine learning model is as follows:

\begin{equation}
\label{shap_eq1}
\phi_{i}=\sum_{S\subseteq N \setminus \left\{ i \right\}}\frac{\left| S \right|! \left( \left| N \right|-\left| S \right|-1 \right)!}{\left| N \right|!}\left[ f\left( S\cup \left\{ i \right\} \right)-f\left( S \right) \right].
\end{equation}

\section{Results and discussion}  \label{sec4}

\subsection{\gls{soh} prediction}
This work aims to provide insight into the \gls{soh} prediction research field by comparing widely used methods for processing data collected through \gls{bms} and generating prediction models for \gls{soh} prediction. In particular, many works are using \gls{pcc} for feature engineering \cite{wen2022soh, tang2023prediction, qu2019neural}. In fact, we believe that global sensitivity analysis methods, such as \gls{shap} and Sobol’ indices, can comprehensively analyze the influence of parameters, just like \gls{pcc}. In addition to widely used time-series prediction models such as \gls{rnn}, \gls{lstm}, and Transformer, advanced machine learning models were generated by integrating time-series prediction models with various machine learning algorithms, such as Convolutional Neural Network (CNN), and applied to \gls{soh} prediction (Table \ref{table_method}). Here, the DLinear, which has strengths in learning patterns of time-series data, is applied, and its accuracy is compared with that of \gls{lstm} and Transformer. According to Zeng et al. \cite{zeng2023dlinear}, the performance of the DLinear is far superior to that of the Transformer model in capturing temporal dynamic patterns between the set of continuous points. The \gls{soh} of a battery often shows a clear decreasing trend even when considering the capacity increase phenomenon in early cycles. Therefore, we believe that DLinear may be a suitable method for \gls{soh} prediction. However, for the fundamental performance comparison of the three algorithms, advanced models through integration with other \gls{ml} algorithms are not considered.

\begin{table}[ht!]
\centering
\caption{Literature on advanced \gls{ml} models for \gls{soh} prediction}
\resizebox{5in}{!}{%
\begin{tabular}{p{5cm}p{8cm}p{8cm}}
\hline
Reference & Methodology & Hyperparameter optimization \\ \hline
    Fu et al. \cite{fu2024lithium}      &  Variational mode decomposition(VMD)-Permutation entropy(PE)-Improved dung beetle optimization (IDBO)-Temporal convolutional network(TCN) & IDBO  \\
     Xia et al. \cite{xia2024soh}     &     Bi-directional gated recurrent unit (BiGRU) & Particle swarm optimization (PSO)       \\
     He et al. \cite{he2024soh}     &    Gaussian process regression (GPR) & Improved PSO with mutation factor and self-adaptive weight adjustment       \\
   Zhou et al. \cite{zhou2024state}       &    Convolutional bi-directional \gls{lstm} (CNN-Bi-LSTM) neural network & Sparrow search algorithm (SSA)        \\
  Zhang et al. \cite{zhang2024soh}        &    Mixed Kernel function relevance vector regression (MKRVR)        &    Differential evolution (DE) with gray wolf optimizer (GWO)     \\
   Zheng et al. \cite{zheng2025joint}       & GRU-TCN-Support vector machine (SVR) (GRU-TCN: Meta learner, SVR: Base learner)   &   Whale optimization algorithm (WOA)    \\
    Tang et al. \cite{tang2025deep}      &   CNN-BiGRU-Attention mechanism (AM)   &  Kepler optimization algorithm (KOA)     \\
     Ma et al. \cite{ma2022novel}       &  \gls{lstm}    &  Differential evolution GWO (DEGWO)     \\
     Zhu et al. \cite{zhu2023state}     &   Bi\gls{lstm}   &       \\ \hline
\end{tabular}
}
\label{table_method}
\end{table}

\begin{figure*}[h!]
    \centering
    \includegraphics[width=0.9\textwidth]{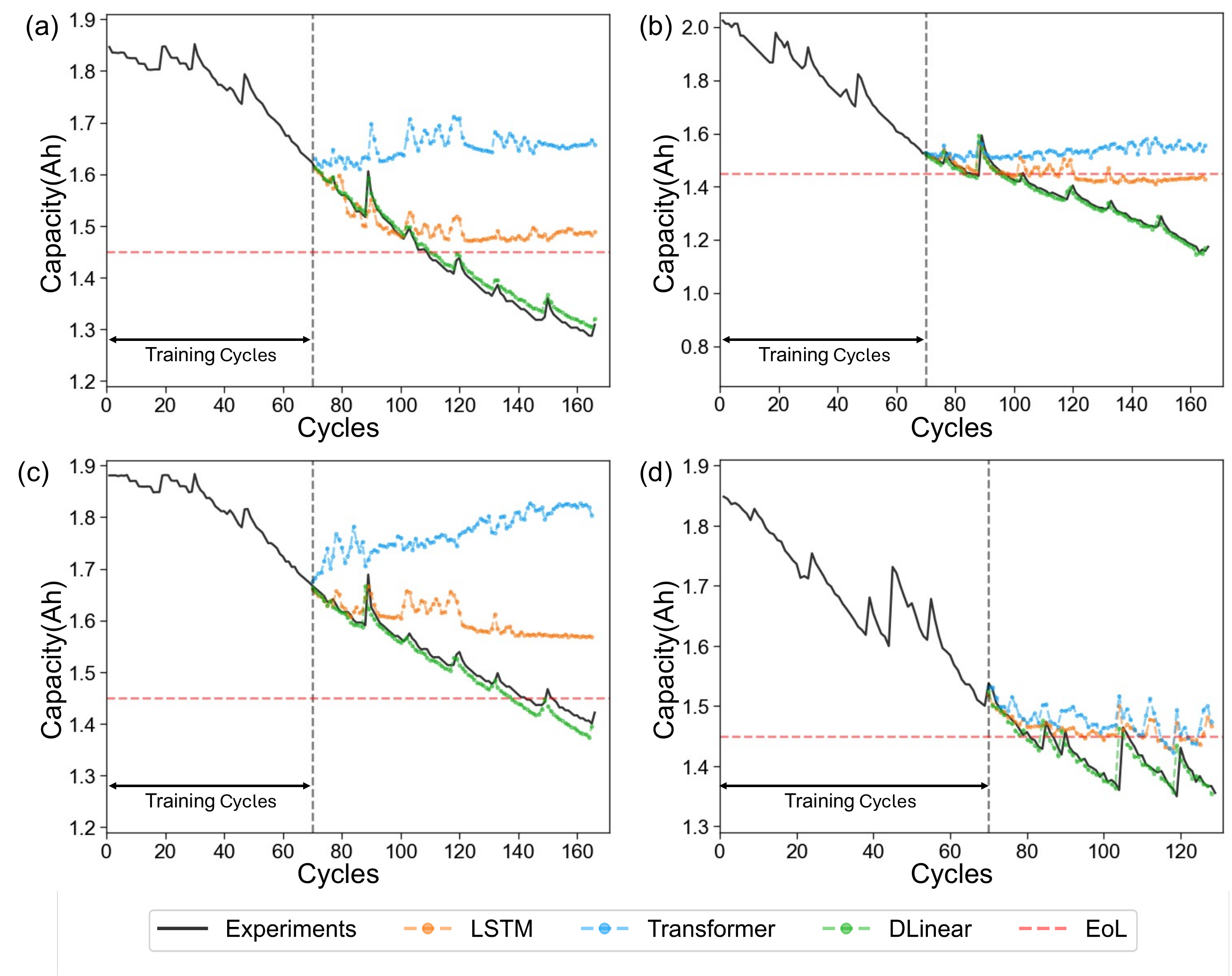} 
    \caption{\gls{soh} prediction results for four cells using all the features specified in the Table \ref{featureTable}: (a) B0005, (b) B0006, (c) B0007, (d) B0018}
\label{fig3}
\end{figure*}

Fig.\ \ref{fig3} shows the prediction results of \gls{soh} for four cells using all the features specified in Table \ref{featureTable}. In order to compare the \gls{soh} prediction accuracy for the four cells under the same criteria, we used up to the 70th cycle as training data. In addition, \gls{soh} of the four cells is defined as the initial capacity of each cell, not the nominal capacity. In this work, the \gls{eol} is defined as reaching 80\% of the initial capacity, and the initial capacity of the four cells and \gls{eol} are different due to \gls{ctcv}. DLinear shows considerable prediction results for all four cells, but LSTM and Transformer do not. In particular, DLinear shows lower \gls{rmse} with fewer training cycles than previous works (Table \ref{table2}). Despite the considerable prediction performance of the DLinear model, the prediction using all 20 features can often cause overfitting or increase the training time, which can be the main cause.

\begin{table}[ht!]
\centering
\caption{Comparison of RMSE across NASA battery cells with previous works}
\resizebox{5in}{!}{%
\begin{tabular}{ccccccc}
\hline
\multirow{2}{*}{Battery} & \multicolumn{2}{c}{Ref.\ \cite{park2020lstm}} & \multicolumn{2}{c}{Ref.\ \cite{kai2024indirect}} & \multicolumn{2}{c}{Our work} \\ \cline{2-7} 
                         & Training cycles  & RMSE & Training cycles  & RMSE & Training cycles    & RMSE    \\ \hline
B0005           & 110        &    0.0168  &          80   &   0.0147   &  70          &   0.0055     \\
B0006           & 90         &    0.0152  &          80   &   0.0196   &  70          &   0.0055      \\
B0007           & 140        &    0.0085  &          80   &   0.0110   &  70          &   0.0084      \\ \hline
\end{tabular}
}
\label{table2}
\end{table}

\subsection{Feature engineering for \gls{soh} prediction}

\begin{figure*}[ht!]
    \centering
    \includegraphics[width=5.0in]{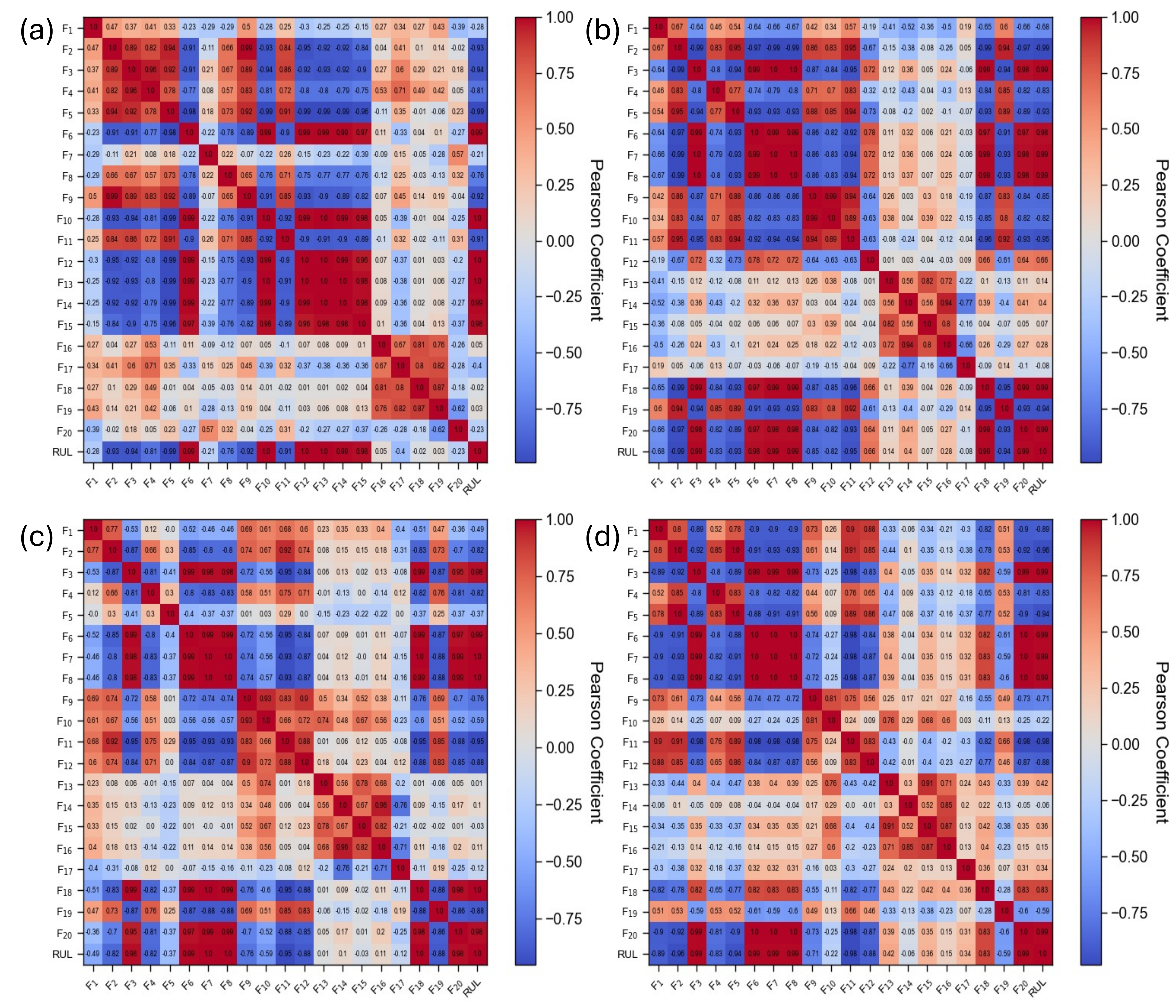} 
    \caption{\gls{pcc} of all features and \gls{soh} for four battery cells: (a) B0005 cell, (b) B0006 cell, (c) B0007 cell, (d) B0018 cell}
\label{fig4}
\end{figure*}

Fig. \ref{fig4} presents the results of \gls{pcc} on 20 features for four cells. The three most sensitive features for each cell, as identified by the \gls{pcc}, are:

\begin{itemize}
\item B0005: F$_{10}$ (Average temperature in discharging), F$_{12}$ (Skewness of temperature in discharging), F$_{13}$ (Minimum temperature in discharging)
\item B0006: F$_{3}$ (Median of loaded voltage in discharging), F$_{18}$ (CC charging time), F$_{20}$ (Total discharging time)
\item B0007: F$_{7}$ (Slope of discharge voltage curve, 50s to 500s), F$_{8}$(Slope of discharge voltage curve, 50s to 1000s), F$_{18}$(CC charging time)
\item B0018: F$_{7}$ (Slope of discharge voltage curve, 50s to 500s), F$_{8}$ (Slope of discharge voltage curve, 50s to 1000s), F$_{20}$ (Total discharging time)
\end{itemize}

In the B0005 cell, the three features (F$_{10}$, F$_{12}$, F$_{13}$) with large values of \gls{pcc} are all temperature-related features. On the other hand, the remaining three cells include voltage-related features in the discharge phase and time-related features in the charge or discharge phase as key features. The reason why the correlations with \gls{soh} are different, even though all four cells are cycled by the same charge and discharge protocol, can be explained by \gls{ctcv}. In particular, it is noteworthy that no feature is identified as a key feature for all four cells.

\begin{figure*}[ht!]
    \centering
    \includegraphics[width=6.5in]{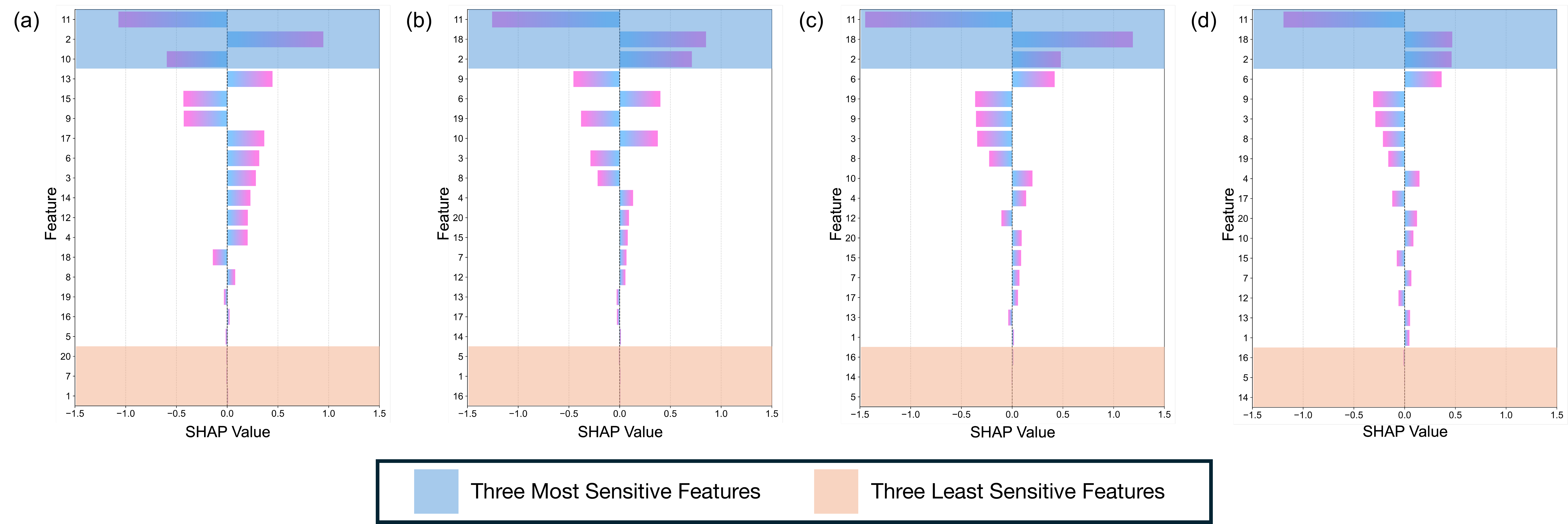} 
    \vspace{-0.1cm}
    \caption{SHAP score of all parameters and \gls{soh} for four battery cells: (a) B0005 cell, (b) B0006 cell, (c) B0007 cell, (d) B0018 cell}
\label{fig5}
\end{figure*}

Fig.\ \ref{fig5} shows the results of global sensitivity analysis based on \gls{shap} for 20 features. The three most sensitive features for each cell, as identified by the \gls{shap}, are:

\begin{itemize}
\item B0005: F$_{2}$ (Variance of measured voltage in discharging), F$_{10}$ (Average temperature in discharging), F$_{11}$ (Variance of temperature in discharging)
\item B0006, B0007, and B0018: F$_{2}$ (Variance of measured voltage in discharging), F$_{11}$ (Variance of temperature in discharging), F$_{18}$ (CC charging time)
\end{itemize}

It is noteworthy that B0006, B0007, and B0018 cells are all explained by the same features in the \gls{shap} results. In addition, B0005 also includes F$_{2}$ and F$_{11}$ among the three features as key features. Unlike \gls{pcc}, the results in \gls{shap} show that there is a common feature subset across the four cells, which means that \gls{shap} can provide feature selection results that successfully reflect \gls{ctcv}.

\begin{figure*}[h!]
    \centering
    \includegraphics[width=\textwidth]{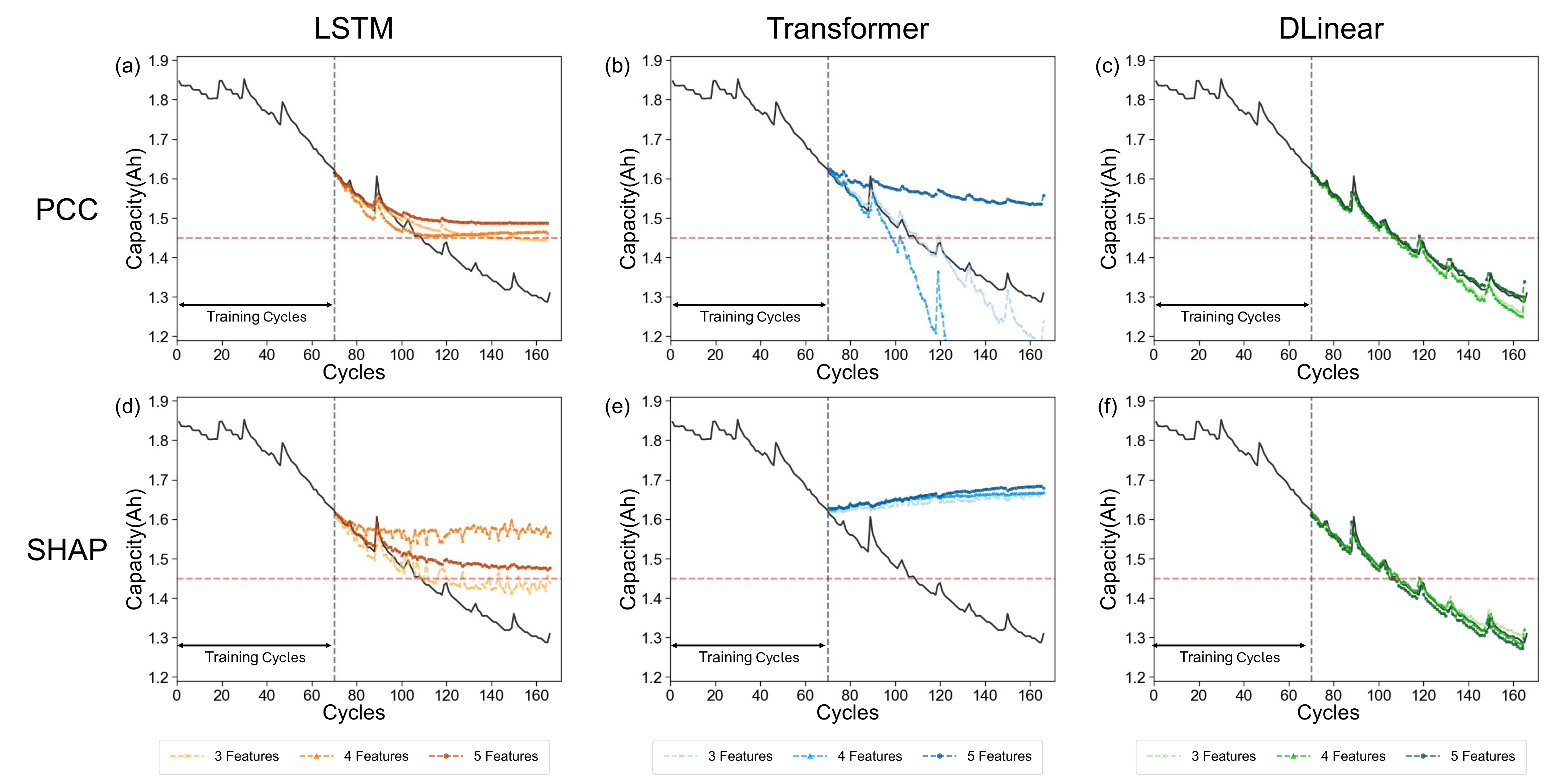}
    \caption{Comparison of results by time-series prediction model according to feature selection method and number of key features: (a) PCC-LSTM, (b) SHAP-LSTM, (c) PCC-Transformer, (d) SHAP-Transformer, (e) PCC-DLinear, and (f) SHAP-DLinear.}
\label{fig6}
\end{figure*}

Fig.\ \ref{fig6} shows the results applied to B0005 cell using 70 cycles for training to compare the accuracy of combinations of two feature selection methods and three time-series prediction models. We also investigated the effect of the number of features by training the model with three, four, and five features. This analysis aimed to validate our hypothesis that simply increasing the feature count does not guarantee better performance. For example, Fig.\ \ref{fig6}(a) and  Fig.\ \ref{fig6}(d) present the results of training \gls{lstm} using features extracted from \gls{pcc} and \gls{shap}, respectively.

Firstly, for \gls{lstm}, the results are almost similar between \gls{pcc} and \gls{shap}. Also, increasing the number of features does not necessarily increase prediction accuracy. A common phenomenon in all \gls{lstm} applications is that, after several cycles in the prediction phase, a flat \gls{soh} profile emerges, indicating that the \gls{lstm} model may encounter gradient vanishing issues.

Secondly, for the Transformer model, \gls{pcc} shows significantly improved predictive performance compared to \gls{lstm}. In the case of \gls{pcc}, the prediction accuracy decreases substantially when four or five features are used, compared to using just three features. This suggests that including features with lower importance ranks may lead to a decrease in accuracy due to increased complexity rather than a model improvement. Furthermore, when using \gls{shap}, it can be explained that the Transformer model does not effectively capture the patterns required for time-series prediction.

Thirdly, the DLinear model exhibits considerable predictive performance with both \gls{pcc} and \gls{shap}. Moreover, there is no significant improvement in accuracy as the number of features increases, with either feature selection method. The features regarding minimum temperature (F$_{13}$, F$_{15}$), identified as the 4$^\text{th}$ and 5$^\text{th}$ most important by the \gls{shap} analysis, are likely to contain already information in the existing temperature-related features (F$_{10}$, F$_{11}$). For example, F$_{10}$ and F$_{13}$ not only show visually similar patterns, but also show significant multicollinearity with a correlation coefficient of 0.5259 (Fig.\ \ref{figA:Temp_features}). In addition, the average temperature is a key indicator because it also reflects the aging effect due to the increase in internal resistance. Therefore, the model accuracy did not improve significantly after including these features due to the information redundancy. In addition, when the number of features is reduced to two, the \gls{soh} prediction performance decreases significantly (Fig.\ \ref{figA:2features}). In summary, among the three time-series prediction models, DLinear is suitable for \gls{soh} prediction, and in our work, it can achieve high accuracy with only three key features.

\begin{figure*}[h!]
    \centering
    \includegraphics[width=\textwidth]{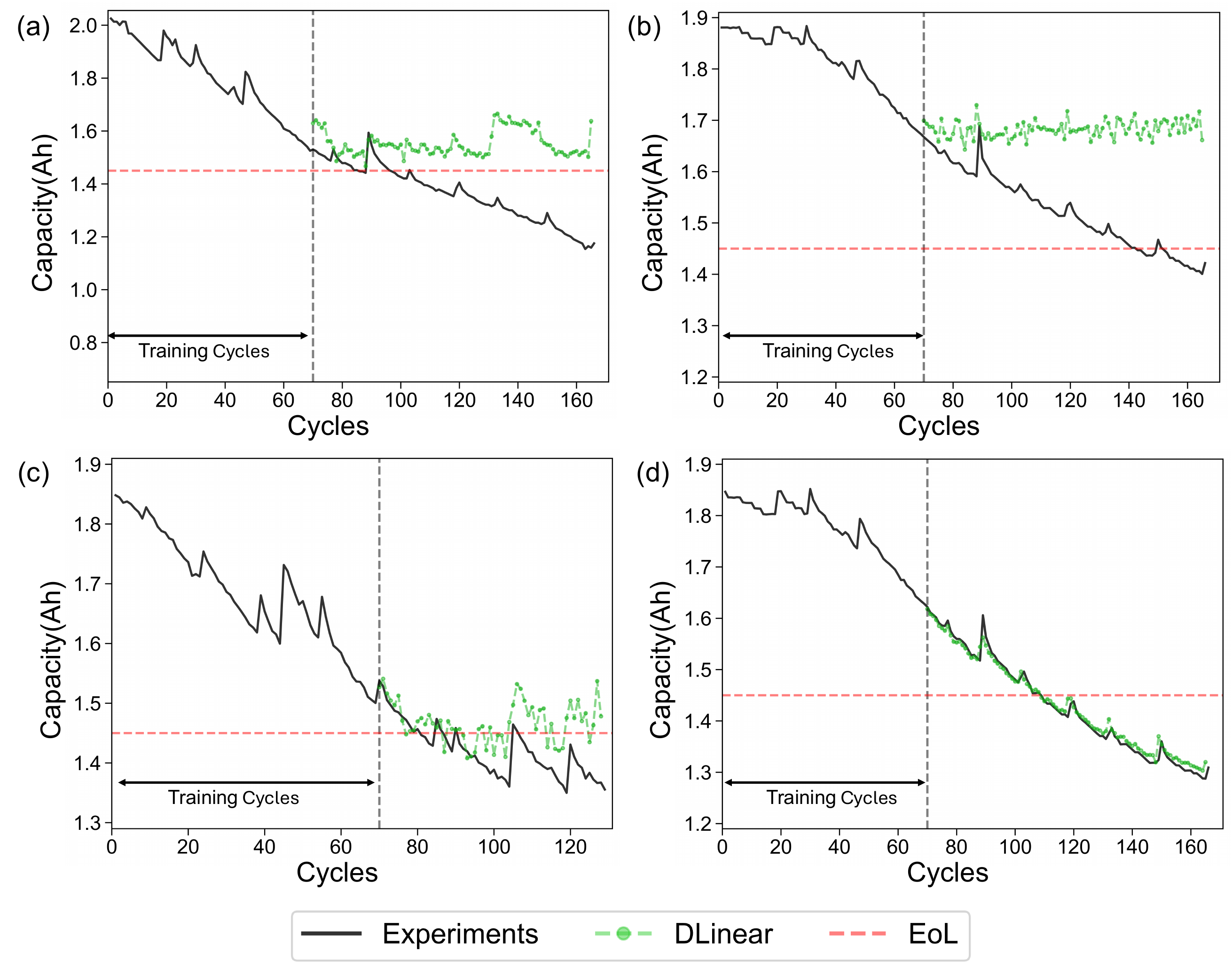}
    \caption{The \gls{soh} prediction of applying the key feature set of B0005 identified by PCC to other cells ((a): B0006, (b): B0007, (c): B0018) and (d) the \gls{soh} prediction of applying the key feature set of B0006 identified by SHAP to B0005}
\label{fig7}
\end{figure*}

\subsection{Robust feature engineering to account for \gls{ctcv}}

In practical \gls{bms} applications in \gls{ev} industry, battery modules or packs consist of dozens or hundreds of cells. Therefore, performing feature engineering individually for all these cells is practically impossible. Selecting a common set of key features for cells with the same battery chemistry will significantly contribute to practical \gls{soh} prediction. Fig.\ \ref{fig7}(a-c) shows the results of applying the key features (i.e., F$_{2}$, F$_{10}$, F$_{11}$) identified by \gls{pcc} on the B0005 cell to \gls{soh} prediction for B0006, B0007, and B0018 cell. The prediction accuracy decreases significantly when compared to the accuracy achieved by training with the unique set of key features identified by applying \gls{pcc} to each individual cell (Fig.\ \ref{fig6}(a-c)). On the other hand, when applying \gls{shap} to identify the key features (i.e., F$_{2}$, F$_{11}$, F$_{18}$) on B0006, B0007, and B0018 cell and then applying them to B0005 cell, the prediction accuracy is significantly higher (Fig.\ \ref{fig7}(d)). One possible interpretation of these results is that using the same key features can yield acceptable performance for cells with the same battery chemistry. This implies that feature selection using \gls{shap} offers robust feature selection results that can account for \gls{ctcv}.

\begin{figure*}[h!]
    \centering
    \includegraphics[width=\textwidth]{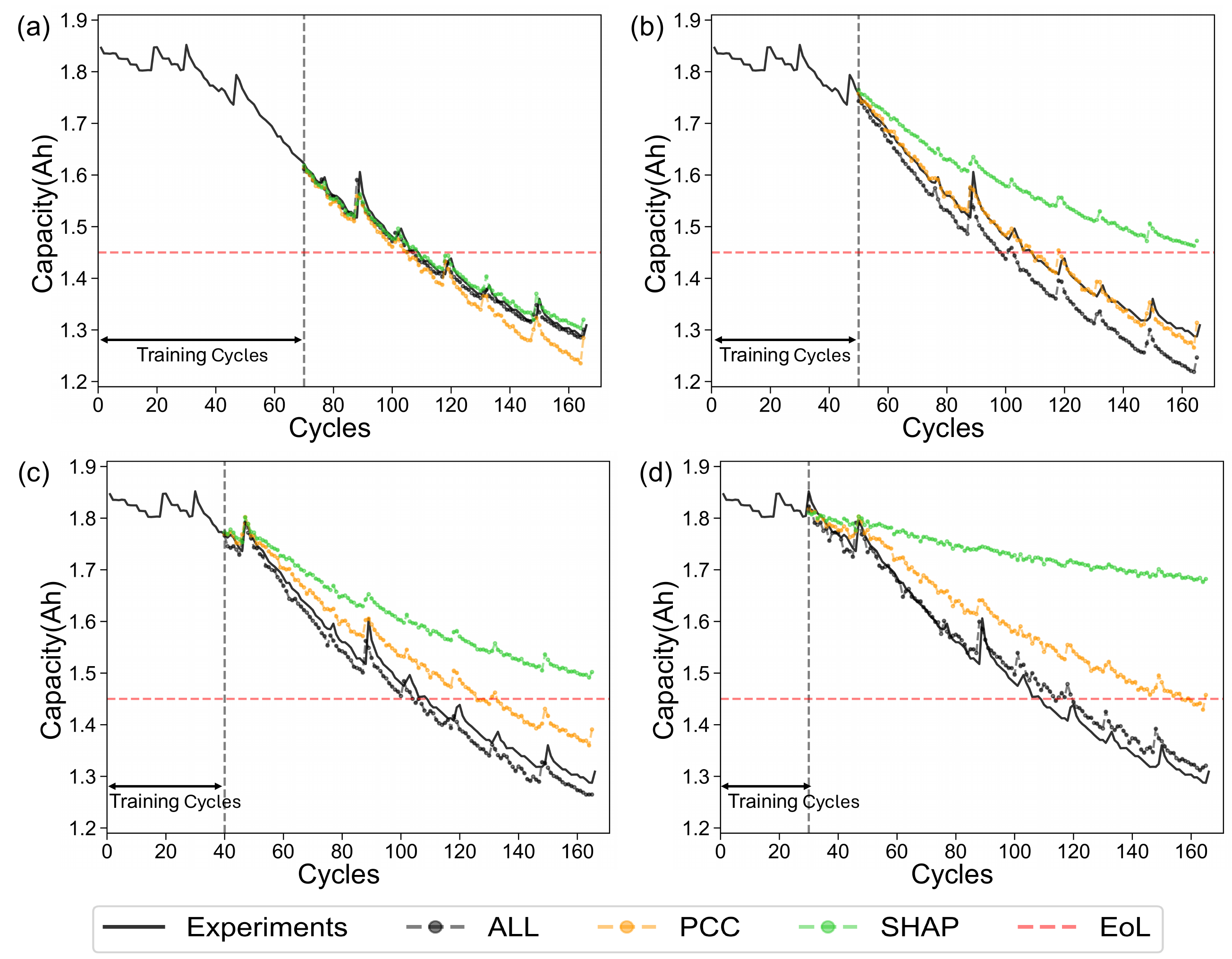}
    \caption{Comparison of \gls{soh} predictions between feature selection methods according to training cycle: (a) 70 cycle, (b) 50 cycle, (c) 40 cycle, (d) 30 cycle.}
\label{fig8}
\end{figure*}

Despite the high robustness of the \gls{shap}-based approach, the rapid decrease in prediction accuracy with reduced training cycles suggests the need for a feature selection method that maximizes both robustness and performance. When training for 70 cycles, all the cases using all features, \gls{pcc}-based selection cases, and \gls{shap}-based selection cases have fairly high prediction accuracies. When considering all features, there is no significant difference between the real \gls{soh} and predicted \gls{soh} even as the training cycles are reduced to 50, 40, and 30. However, when the number of training cycles decreases, the accuracy rapidly declines with \gls{shap}. However, with \gls{pcc}, predicted \gls{soh} remain very close to the real \gls{soh} with 50 cycles for training and still effectively track the trend of the real \gls{soh} when training cycles are reduced to 40 or 30. This suggests that \gls{pcc} might be more appropriate for predicting the \gls{soh} of individual cells and during extremely early cycles, while \gls{shap} could be more suitable for predictions at the module or pack scale.

\section{Conclusion} \label{sec5}

Accurate and efficient prediction of \gls{soh} is essential for realizing condition-based maintenance in \gls{ev} applications. In this work, we proposed an explainable \gls{soh} prediction framework combining feature engineering and the DLinear model. Leveraging public battery aging datasets, we demonstrated that DLinear outperforms conventional \gls{lstm} and Transformer models in both prediction accuracy and computational efficiency, even when trained with limited cycling data.
To address the challenge of \gls{ctcv}, we employed Pearson correlation and \gls{shap}-based sensitivity analyses for feature selection. While \gls{pcc} offered strong performance for individual cells, \gls{shap} consistently identified key features applicable across multiple cells, enabling generalized \gls{soh} prediction for \gls{ev} battery modules. This is especially critical in large-scale battery systems, where individual cell-level modeling is impractical.
The proposed framework offers a scalable, interpretable, and low-complexity solution with significant practical advantages for \gls{ev} \gls{bms}. The feature engineering process facilitates the generation of a dataset that is optimized for \gls{ml} model training by enabling a substantial reduction in the number of features a \gls{bms} needs to measure and store. Furthermore, the use of \gls{shap} ensures that key factors like \gls{ctcv} are considered, while the DLinear model's linear structure provides a substantially lower computational burden than more complex architectures like Transformer and \gls{lstm}. It enables early and reliable detection of battery degradation trends, contributing to safer operation, reduced maintenance costs, and improved battery utilization in electric vehicles.
Future work will focus on integrating real-world \gls{ev} cycling data, refining model robustness under diverse usage conditions, and extending the framework for online learning in embedded \gls{bms} platforms.

\section*{Code Availability}
The code used in this work will be available after publication.


\clearpage

\appendix \label{sec:appendix}
\setcounter{table}{0}
\renewcommand{\thetable}{A-\Roman{table}}

\appendix 
\label{sec:appendix}
\setcounter{table}{0}
\renewcommand{\thetable}{A-\Roman{table}}
\setcounter{equation}{0}
\renewcommand{\theequation}{A-\arabic{equation}}
\setcounter{figure}{0}
\renewcommand{\thefigure}{A-\arabic{figure}}

\section{Feature selection of B0005 cell} \label{Appendix}

\begin{figure}[h!]
\centering
\includegraphics[width=0.9\textwidth]{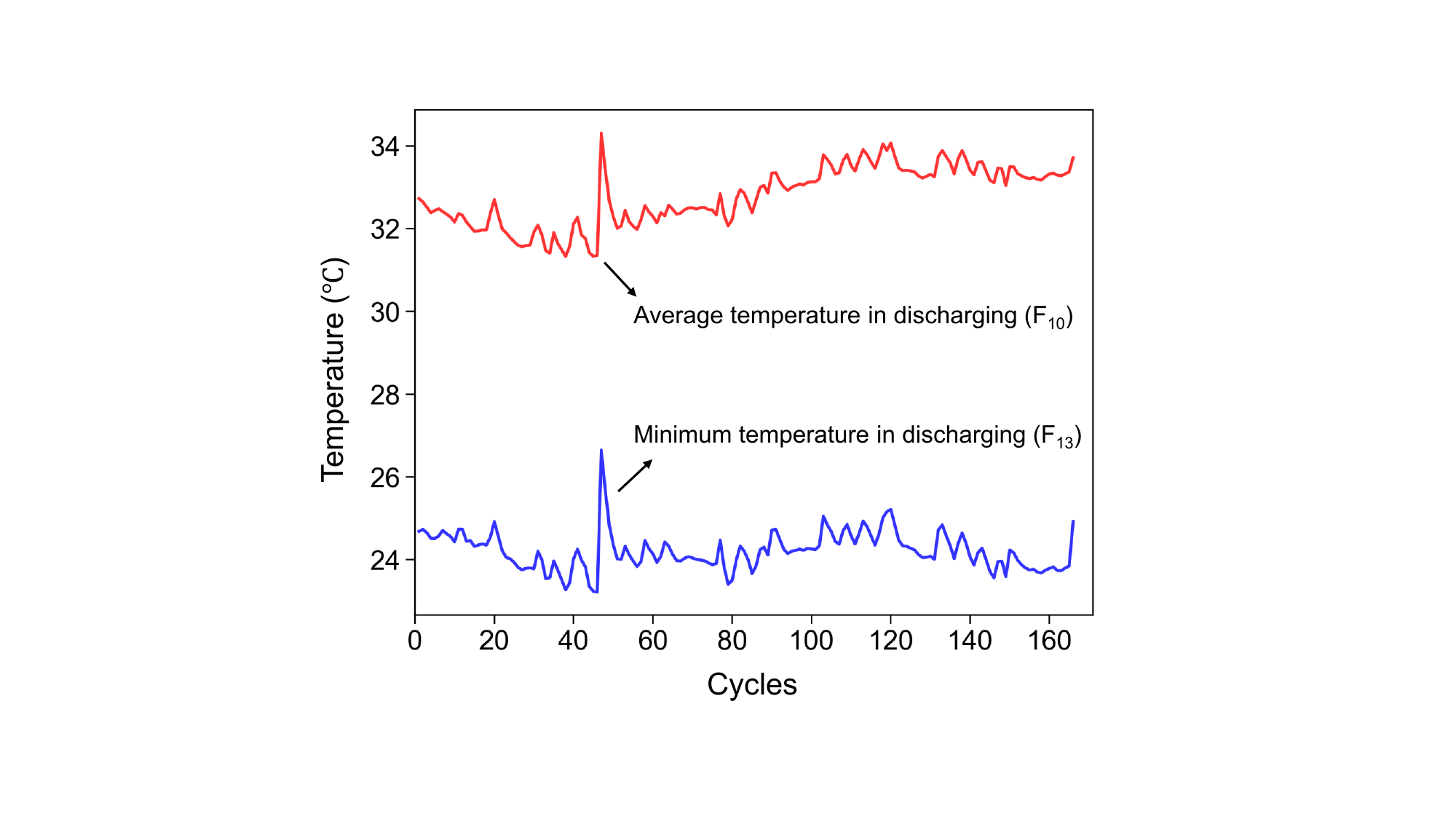}
\vspace{-1.5cm}
\caption{Two temperature parameters (F$_{10}$ and F$_{13}$) of B0005 cell}
\label{figA:Temp_features}
\end{figure}

\begin{figure}[h!]
\centering
\includegraphics[width=0.5\textwidth]{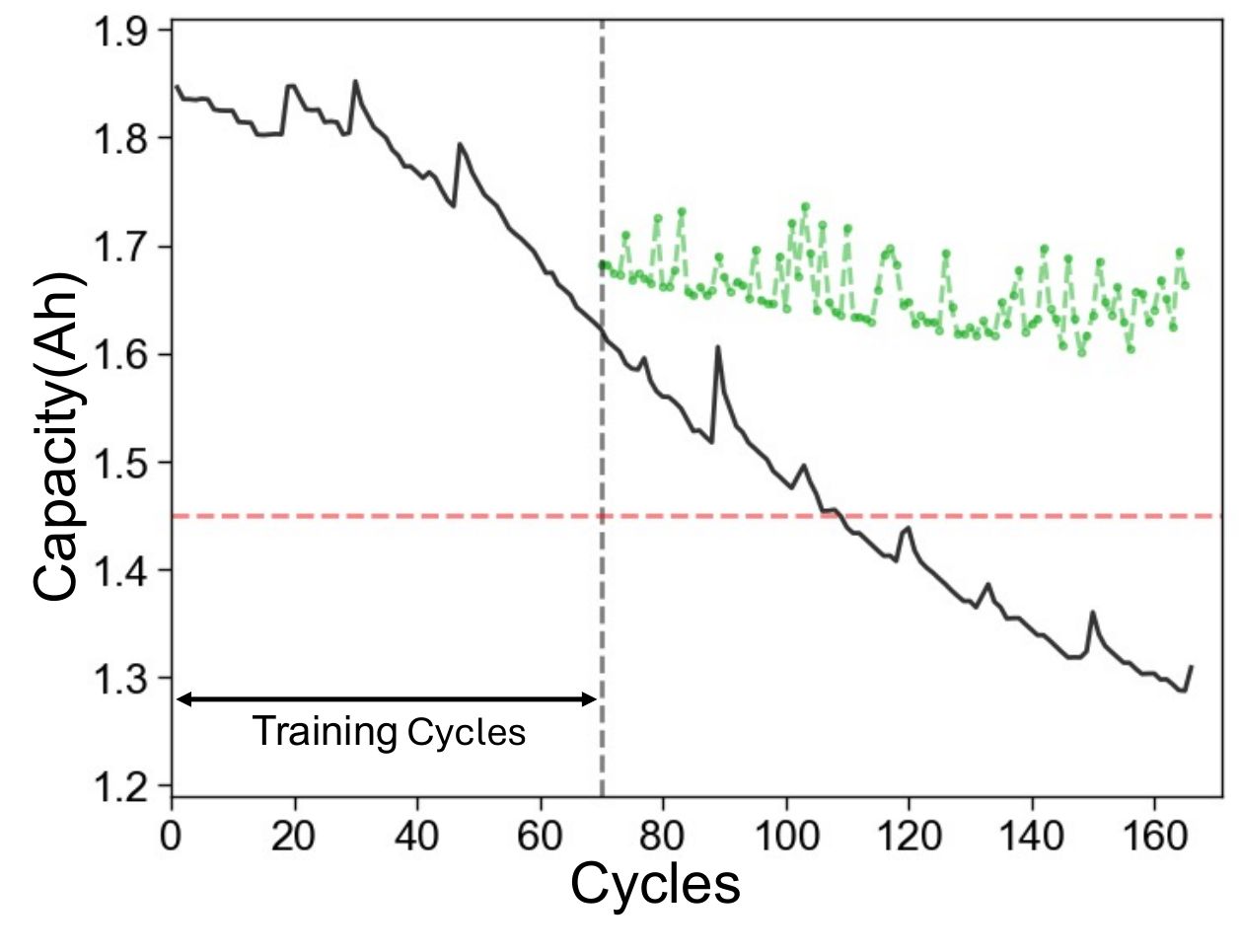}
\vspace{-0.5cm}
\caption{DLinear-based prediction results of B0005 cell using the two most sensitive parameters identified by \gls{shap}}
\label{figA:2features}
\end{figure}

\clearpage

\bibliography{apssamp}

\end{document}